\documentclass[conference]{IEEEtran}
\usepackage{cite}
\usepackage{hyperref}
\usepackage{comment}

\ifCLASSINFOpdf
  \usepackage[pdftex]{graphicx}
\else
  \usepackage[dvips]{graphicx}
\fi
\ifCLASSOPTIONcompsoc
  \usepackage[caption=false,font=normalsize,labelfont=sf,textfont=sf]{subfig}
\else
  \usepackage[caption=false,font=footnotesize]{subfig}
\fi
\begin{document}
%
\title{Quantum Computing in Intelligent Transportation Systems: A Survey}

\author{
    \IEEEauthorblockA{
        \begin{tabular}{cccc}
            Yifan Zhuang & Talha Azfar & Yinhai Wang & Wei Sun\\
            Google & Rensselaer Polytechnic Institute & University of Washington & AIWaysion Inc.\\
            Mountain View, CA, USA & Troy, NY, USA & Seattle, WA, USA & Seattle, WA, USA\\
            zhuangyifan@google.com & azfart@rpi.edu & yinhai@uw.edu & wsun@aiwaysion.com
        \end{tabular}
        \vspace{0.5cm} \\
        \begin{tabular}{ccc}
            Xiaokun (Cara) Wang & Qianwen (Vivian) Guo & Ruimin Ke* \\
            Rensselaer Polytechnic Institute & Florida State University & Rensselaer Polytechnic Institute \\
            Troy, NY, USA & Tallahassee, FL, USA & Troy, NY, USA \\
            wangx18@rpi.edu & qguo@fsu.edu & ker@rpi.edu
        \end{tabular}
    }
}

%


\maketitle

\begin{abstract}
Quantum computing, a field utilizing the principles of quantum mechanics, promises great advancements across various industries. This survey paper is focused on the burgeoning intersection of quantum computing and intelligent transportation systems, exploring its potential to transform areas such as traffic optimization, logistics, routing, and autonomous vehicles. By examining current research efforts, challenges, and future directions, this survey aims to provide a comprehensive overview of how quantum computing could affect the future of transportation.
\end{abstract}

%
\IEEEpeerreviewmaketitle

\section{Introduction}
The rapid advancement of quantum computing technology has opened up new horizons for solving complex problems across various domains, including intelligent transportation systems (ITS). ITS encompasses a wide range of technologies and applications to improve transportation networks' efficiency, safety, and sustainability \cite{zhu2018big}. However, traditional computational approaches often struggle to handle the massive amounts of data and intricate optimization problems inherent in ITS. With its unique capabilities in parallel processing, optimization, and machine learning, quantum computing offers new ways to manage and optimize transportation systems \cite{wang2021shaping}.

This potential has spurred significant research interest in applying quantum computing to various facets of ITS. Quantum algorithms have shown promise in specific areas such as traffic flow optimization, intelligent routing, autonomous driving, and traffic prediction. These algorithms can potentially outperform classical counterparts in speed, efficiency, and accuracy by leveraging quantum principles like superposition and entanglement. Moreover, quantum machine learning techniques might help uncover hidden patterns and correlations within vast traffic datasets, leading to more informed decision-making and predictive capabilities.

However, the integration of quantum computing into ITS poses certain challenges. The development of scalable and fault-tolerant quantum hardware, a task of immense complexity, remains a significant hurdle. Additionally, designing quantum algorithms tailored for specific ITS applications requires expertise in both quantum computing and transportation engineering, highlighting the importance of interdisciplinary collaboration. Despite these challenges, the potential benefits of quantum computing for ITS are immense, making the journey towards integration all the more worthwhile and significant.

This survey paper aims to provide a comprehensive and engaging overview of the current state of quantum computing in ITS, which is still in an early stage. The following content will explore the theoretical foundations of quantum computing, review recent advancements in quantum algorithms relevant to ITS, and discuss the challenges and opportunities associated with this emerging field. By examining case studies and real-world applications, this paper will shed light on the potential impact of quantum computing on the future of transportation systems.

\section{Universal Quantum Computing}
Before reviewing quantum computing applications in ITS, it is necessary to introduce prior knowledge of quantum computing (Section \ref{subsec:prior}) and quantum machine learning (Section \ref{subsec:qml}) as it is an emerging technology and not familiar to most researchers.

\subsection{Prior Knowledge} \label{subsec:prior}
Quantum computing is a revolutionary computing paradigm that makes use of the principles of quantum mechanics to solve complex problems that are practically impossible for classical computers \cite{abdelgaber2020overview}. Unlike classical computers that use bits (0 or 1), quantum computers use qubits existing in a superposition of states (0 and 1 simultaneously), allowing quantum computers to perform many calculations at once. Before diving into comparison with classical computers and quantum computing applications, it is necessary to briefly introduce basic quantum computing knowledge and algorithms.

Gate-based quantum computing uses quantum gates (analogous to logic gates in classical computers) to manipulate qubits \cite{michielsen2017benchmarking}. As mentioned above, n bits can represent one of $2^n$ numbers in classical computers. An appropriate state of n qubits can represent all $2^n$ numbers simultaneously in quantum superposition. When the quantum gates operate on qubits, any operation is applied to all states simultaneously, resulting in an exponential speedup. However, reading the result of such a quantum-parallel operation directly is impossible but relies on finding the state probability $P(x)$ based on the Born rule. The computation formula is shown in Equation \ref{equ_born_prob}.
\begin{equation} \label{equ_born_prob}
    P(x)=A(x)^* A(x)=|A(x)|^2=\langle x|x \rangle,
\end{equation}
where $A(x)$ indicates the amplitude of state x and $*$ is the complex conjugate operation. The dot product $\langle x|x \rangle$ shows the same quantity expressed in Dirac notation. Currently, quantum computing relies on manipulating the phase of quantum amplitudes to effect the desired result when amplitudes are combined. Thus, quantum amplitudes could only be manipulated via unitary transformations, excluding some taken-for-granted operations in classical computers, e.g., copying state and deleting state information.

Two representative algorithms based on gated-based quantum computing are Shor's algorithm designed for finding the prime factors of an integer \cite{shor1994algorithms}, and Grover's algorithm targeting for unstructured search \cite{grover1996fast}. Shor's algorithm leverages the quantum Fourier transform and other quantum properties to find the period of a function, which can then be used to factor the number. Grover's algorithm uses amplitude amplification to increase the probability of finding the correct item in the database. In terms of algorithm complexity, Shor's algorithm owns the complexity $O(\log(n))$ while the traditional brute force search has the complexity $O(\sqrt{n})$. Grover's algorithm could reach $O(\sqrt{n})$ compared to classical algorithms' $O(n)$. 

Although gated-based quantum computing provides a universal model, it requires precise control of qubits, which are highly susceptible to errors. Quantum Annealing (QA) is a significant technique for quantum optimization problems  \cite{finnila1994quantum}, which leverages a quantum phenomenon to find the lowest energy state of a system. This energy state corresponds to the optimal solution of a specific problem. Annealing techniques are a practical realization of the theoretical adiabatic quantum computation model. This approach, similar to classical simulated annealing, uses a quantum wave function for approximate optimization \cite{farhi2000quantum}. The state of qubits can be designed to represent candidate solutions, with lower energy levels signifying better solutions. QA uses Hamiltonian matrix to simulate a cooling process, starting with high kinetic energy. It allows access to all configurations, then gradually reduces it to favor low-energy, so-called good solutions. While infinite cooling time guarantees optimality, practical limitations lead to approximate results. Quantum annealing generally offers quadratic improvement over classical annealing, with potential for exponential improvement in specific cases \cite{roland2002quantum}. In fact this speedup has been demonstrated in applications such as transport network design problems \cite{dixit2023quantum}.

Quantum random walks offer a valuable framework for both designing and evaluating the performance of gate-based and quantum annealing algorithms \cite{aharonov1993quantum}. In a classical random walk, a "walker" moves on a line or graph and randomly chooses to move left or right at each step. Over time, the walker's position becomes a probability distribution, often exhibiting diffusive behavior. This behavior extends to the quantum realm by representing the particle as a complex amplitude on each node, enabling superposition and interference, with the final measured position governed by the Born rule \cite{kempe2003quantum}. Instead of a deterministic one-direction move in classical operation, a quantum operation (a unitary gate) is applied to the walker's state, resulting in movement in both directions simultaneously with different probabilities.

\subsection{Quantum Machine Learning} \label{subsec:qml}
Conventional machine learning  algorithms have been significantly benefiting from classical computers equipped with GPU/TPU for fast processing, e.g., vision-based traffic monitoring \cite{dilek2023computer}, traffic prediction \cite{yuan2021survey}, and traffic management \cite{patil2022applications}. Compared with classical computers, quantum computers could further bring exponential speedup for specific problems \cite{biamonte2017quantum, marella2020introduction, batra2021quantum} by exploiting quantum phenomena like superposition and entanglement, allowing them to explore multiple solutions simultaneously. This part will briefly introduce quantum-version of machine learning algorithms \cite{schuld2015introduction, ramezani2020machine}, including supervised and unsupervised learning.

Artificial Neural Network (ANN), as a fundamental model, is based on aggregation of non-linear functions applied to neurons that are laid out in layers and sequences \cite{narayanan2000quantum}. However, it is challenging to build non-linear activation blocks due to linear nature of quantum mechanics.
Schuld et al., proposed a general procedure for step activation function \cite{schuld2015simulating}.
Cao et al., presented a quantum neuron based on quantum circuit to simulate threshold activation \cite{cao2017quantum}. 
Panella and Martinelli proposed a technique using Boolean functions and nonlinear quantum circuits to transform non-linear data into a linear form~\cite{panella2008neurofuzzy}.
Beer et al., designed a truly quantum analogue of classical neurons, composing quantum feedforward neural networks for universal quantum computation \cite{beer2020training}.

Convolution Neural Network (CNN) is a more complicated ANN model structure. Quantum CNN (QCNN) provides a new solution with CNN using a quantum computing environment, or a direction to improve the performance of an existing learning model \cite{oh2020tutorial}. Kerenidis et al., designed a shallow circuit algorithm for QCNN with non-linear activation and pooling operations \cite{kerenidis2019quantum}. They introduced a new quantum tomography algorithm with l-infinity norm guarantees, and new applications of probabilistic sampling when processing information.
Kerenidis et al., provided a quantum approach for gradient descent computation when the gradient is an affine function, where single step time is considerably smaller than the classical cost \cite{kerenidis2020quantum}.

Generative Adversarial Network (GAN) includes two neural networks, i.e., generator and discriminator, compete against each other in a zero-sum game framework. Wasserstein GAN (WGAN) is a GAN variant that uses Wasserstein distance between outputs from the generator and discriminator, to improve training stability \cite{arjovsky2017wasserstein}.
Chakrabarti et al., firstly designed the quantum WGAN to improve the robustness and the scalability of the adversarial training of quantum generative models on noisy quantum hardware \cite{chakrabarti2019quantum}.
Situ et al., proposed a hybrid quantum GAN with a parameterized quantum circuit generator and a classical neural network discriminator for generating classical discrete distributions. The quantum circuit uses only simple gates available in current devices \cite{situ2020quantum}.

Clustering algorithms, as representative of unsupervised learning, may take polynomial time to solve these problems, while quantum algorithms take logarithmic processing time \cite{anand2022quantum}.
Horn and Gottlieb proposed a novel quantum clustering algorithm inheriting from scale-space clustering and support-vector clustering \cite{horn2001method}. support-vector clustering maps data points to states in Hilbert space, represented by Gaussian wave functions. This allows for assigning weights to specific points, emphasizing them as potential cluster centers.
A{\"\i}meur et al., presented three quantum clustering algorithms, i.e., divisive clustering, k-medians, and c-neighborhood graph construction, utilizing quantization and Grover’s algorithm \cite{aimeur2007quantum}.
Lloyd et al., applied supervised and unsupervised quantum machine learning algorithms in cluster assignment and cluster finding \cite{lloyd2013quantum}. The results obtained exponential speedup by reducing processing time from $O(n)$ to $O(\log(n))$.

\section{Applications in Intelligent Transportation}
Although quantum computing is still in an early stage, it shows great potential in computing efficiency on specific tasks. This benefits from quantum properties like superposition and entanglement, allowing parallel computations. This survey will explore quantum applications in following transportation perspectives -- Section \ref{subsec:management}  traffic management, Section \ref{subsec:routing} vehicle routing, and Section \ref{subsec:autonomous} autonomous driving.

\subsection{Quantum Computing for Traffic Management} \label{subsec:management}
Transportation management encompasses the planning, execution, and optimization of the movement of goods and people \cite{de2017traffic}. It involves overseeing all aspects of the transportation process, from selecting the most suitable modes of transportation and carriers to managing shipping schedules, tracking shipments, and ensuring timely deliveries. Transportation management frequently benefits from automatic optimization \cite{iliopoulou2019combining}. Quantum algorithms could potentially analyze real-time traffic data (e.g., vehicle density, speed, and road conditions), optimize traffic signal timing, and reduce congestion.

\begin{figure}[ht!]
    \centering
    \subfloat[Classical STGCN]
    {\includegraphics[width=0.9\linewidth]{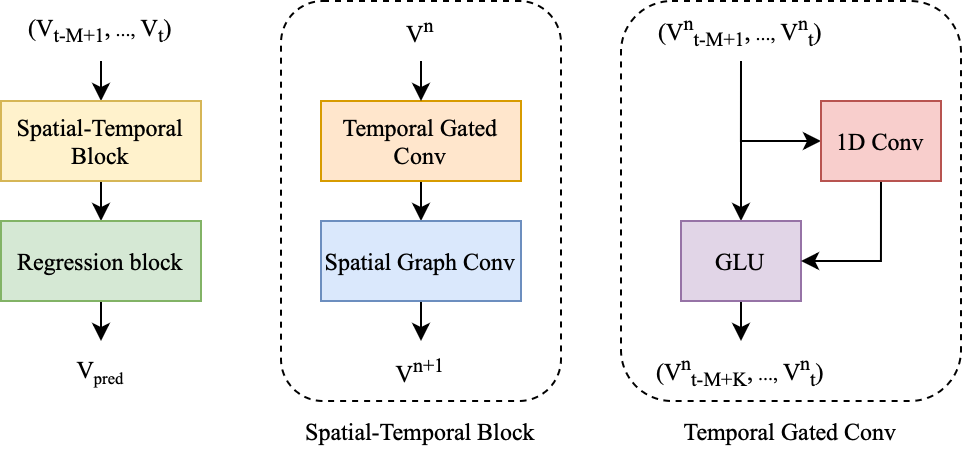}} \\
    \subfloat[Quantum STGCN]
    {\includegraphics[width=0.9\linewidth]{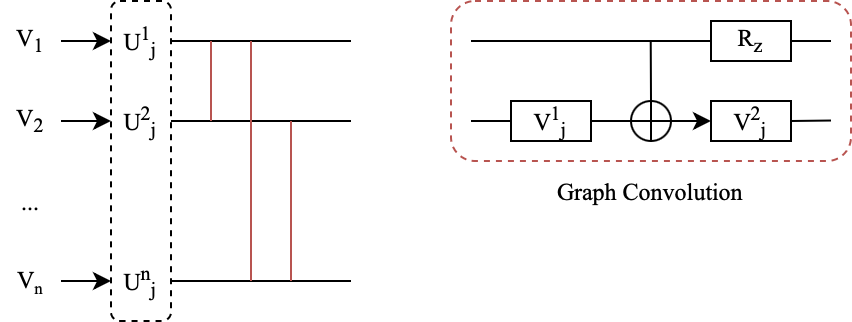}} \\
    \caption[Comparison of Classical and Quantum STGCN]{Comparison of Classical and Quantum STGCN.}\label{fig:comparison_stgcn}
\end{figure}

Zhang et al., introduced a novel traffic flow prediction algorithm using Quantum Genetic Algorithm - Learning Vector Quantization (QGA-LVQ) neural network \cite{zhang2020application}. This approach combines the structural simplicity and effective clustering of LVQ \cite{kohonen2001learning} with the global optimization capabilities of QGA \cite{lahoz2016quantum}. This hybrid model addresses the shortcomings of LVQ, e.g., sensitivity to initial weights and susceptibility to local optimization.
Zhang et al., studied traffic flow forecasting using the Quantum-behaved Particle Swarm Optimization (QPSO) strategy \cite{zhang2021new}. A hybrid approach combining genetic simulated annealing and quantum particle swarm optimization is utilized to determine the optimal initial cluster centers for the radial basis function neural network (RBFNN) prediction model. The RBFNN's function approximation capability is then employed to generate the desired data.
Huang et al. proposed a traffic prediction method leveraging Modified Ensemble Empirical Mode Decomposition (MEEMD) and Quantum Neural Network (QNN) to address the long-range and short-range dependencies of backbone network traffic \cite{huang2018backbone}. The method involves pre-processing traffic data with MEEMD to decompose it into Intrinsic Mode 
Function (IMF) components. Subsequently, the QNN, with its exceptional nonlinear processing and convergence capabilities, predicts each IMF component.
Besides traffic flow prediction, traffic congestion prediction is challenging as well due to dynamic traffic data and complicated traffic networks. Qu et al., presented a quantum Spatial-Temporal Graph Convolution Network (STGCN) by utilizing a closed-form solution from Schrödinger approach to learn the temporal features and constructing a quantum GCN for spatial features  \cite{qu2022temporal}. The comparison between classical STGCN \cite{yu2017spatio} and quantum counterpart is shown in Figure \ref{fig:comparison_stgcn}.

Traffic signal optimization is significant in reducing traffic congestion.
Hu et al., built an enhanced Biham, Middleton and Levine (BML) model to simulate the urban traffic and used QPSO to improve time scheduling \cite{hu2018quantum}. They firstly mapped the real-world road network with different two-way multi-lane roads into the theoretical lattice space of BML. The proposed enhanced BML provides a deeper insight into phase transition and traffic congestion.
Hussain et al., converted the optimal control of time-dependent traffic signals into a Quadratic Unconstrained Binary Optimization (QUBO) \cite{hussain2020optimal}. Considering quantum computer hardware constraints, e.g., limited number of qubits, inter-qubit connectivity and high error rates in D-Wave devices, they provided a hybrid (quantum-classical) approach.
Traffic flow and supply chain logistics optimization has been studied using quantum annealing to solve a QUBO problem in \cite{neukart2017traffic}, which used a D-Wave quantum processing unit to minimize congestion based on the T-drive Beijing road dataset. With larger quantum computers and real-time data, the entire city traffic flows could be optimized in real-time. 
Inoue et al., developed a QA-based method to control large-scale traffic signals \cite{inoue2021traffic}. Initially, they defined a signal optimization problem aimed at minimizing the disparity in traffic flows between two perpendicular directions, which is then transformed into an Ising Hamiltonian, suitable for processing by QA. The new global control method was evaluated in a large city simulation of 50-by-50 intersections, and the results demonstrated the superior performance of the global control method in reducing traffic imbalances across a broad range of parameters.

\subsection{Quantum Computing for Vehicle Routing} \label{subsec:routing}
Transportation routing and scheduling are two interconnected processes within the field of logistics that aim to optimize the movement of goods and services. Routing focuses on determining the most efficient paths or routes for vehicles to reach their destinations \cite{garaix2010vehicle}. It considers various factors, such as distance between stops, traffic conditions, road restrictions, and vehicle capacity. Scheduling involves assigning specific time slots for pickups and deliveries, as well as determining when vehicles should start and end their routes. It takes driver availability, vehicle maintenance schedules, and traffic patterns into accounts \cite{nama2021machine}.

Early works were conducted on the Traveling Salesman Problem (TSP). Moser firstly presented a quantum mechanical solution for the standard TSP and its generalization problems back in 2003 \cite{moser2003quantum}. Martovnak and collaborators then proposed to use QA for the travel salesman problem \cite{martovnak2004quantum}.
Chen et al., experimentally demonstrated applying QA to a simplified TSP by simulating the corresponding Schrödinger equation with a nuclear-magnetic-resonance quantum simulator \cite{chen2011experimental}.
Moylett et al., applied  a quantum backtracking algorithm on a classical algorithm and achieved a quadratic quantum speedup when the degree of each vertex is at most 3 \cite{moylett2017quantum}. Later, quantum computing was extended to more complicated routing problems, e.g., graph traversals \cite{dorn2007quantum} and multi-object routing \cite{osaba2022systematic}. 

Multi-objective routing offers users a variety of routes that balance factors like arrival time and trip duration \cite{zajac2021objectives}. Multi-objective problems lack a single best option because of various individual preferences. For public transport, the ideal route depends on how users prioritize factors like travel time, wait time, arrival time, and other personal considerations.
Zhang et al., developed an adaptive grid Multi-Objective Quantum Evolutionary Algorithm (MOQEA) that considers both customer satisfaction and travel cost \cite{zhang2012multiobjective}. MOQEA uses a modified fuzzy due-time window to represent user satisfaction, encodes chromosomes, and produces multiple non-dominant solution sets. To ensure solution diversity, an adaptive grid adjusts the current grid number based on the previous generation's solution distribution.
Besides general vehicle routing, railway freight routing problem is also a challenging variant. Zhang et al. presented an improved multi-objective QPSO algorithm \cite{zhang2019improved}. Their algorithm achieved superior search capabilities with fewer parameters, improving global search and avoiding local optimization. Compared to other continuous multi-objective swarm intelligence algorithms, their method produced a more accurate Pareto front for the railway freight transportation route design.

Wang et al. introduced a hybrid algorithm vehicle schedule problem called improved Quantum-Inspired Evolutionary Algorithm (IQEA) \cite{wang2012design}. The proposed algorithm combines QEA \cite{han2002quantum} with greedy heuristics to enhance QEA in solving more complex problems, i.e., the Vehicle Routing Problem with Time Windows (VRPTW). IQEA demonstrated to be a faster solution especially for larger-scale instances. In addressing consignment order, IQEA's upper level splits consignment sequences into sub-sequences based on qubit values. The lower level employs greedy heuristics to rebuild these sub-sequences for minimal transportation cost.
Leonidas et al. explored using qubit encoding scheme to reduce number of binary variables in VRPTW \cite{leonidas2023qubit}. Their approach could find the approximate solutions with fewer qubits compared to other quantum algorithms using the full encoding.

\begin{figure}
    \centering
    \includegraphics[width=0.9\linewidth]{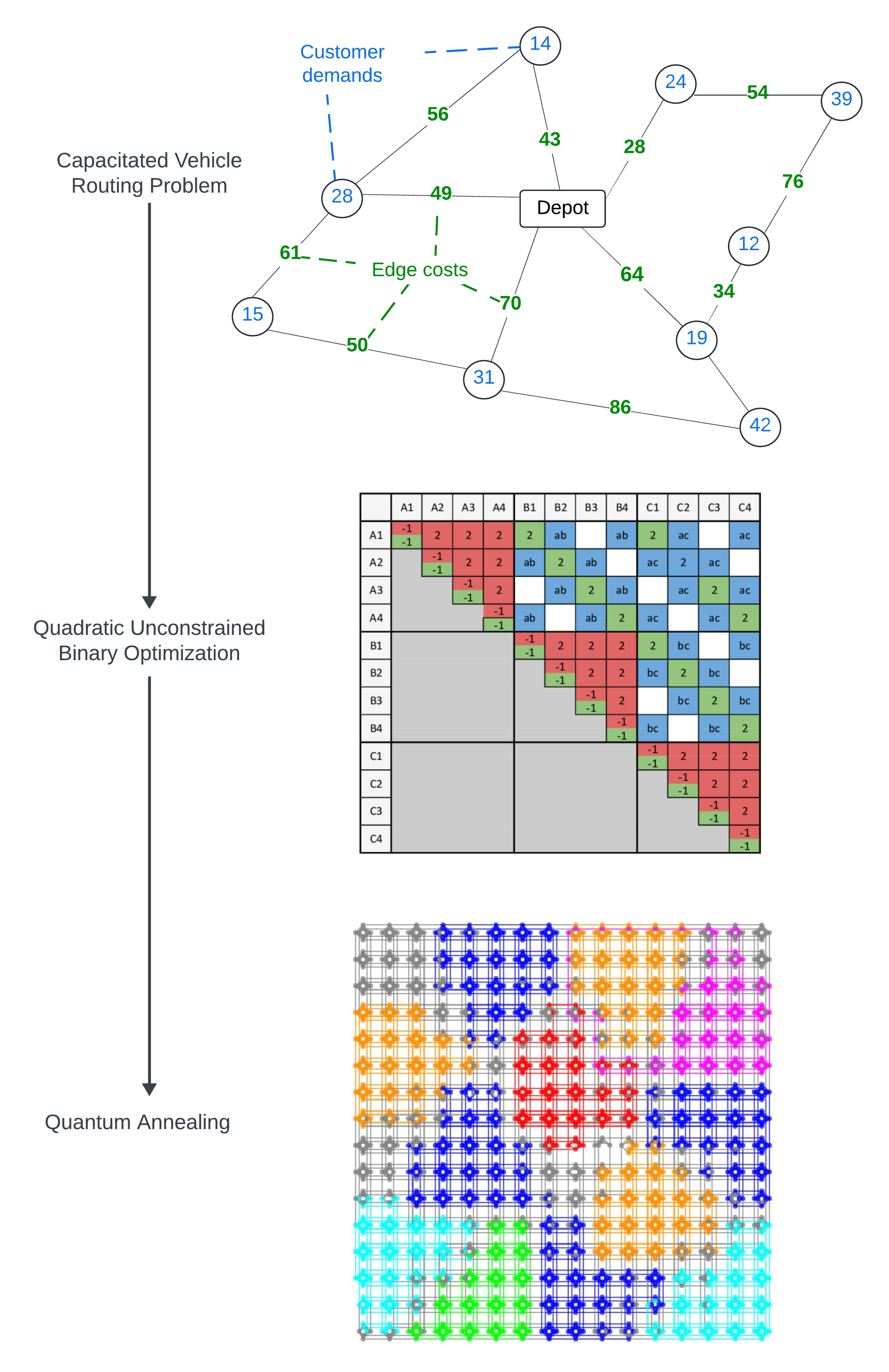}
    \caption{Capacitated vehicle routing problem adapted to quantum annealing by transforming to a quadratic unconstrained binary problem \cite{feld2019hybrid} \cite{pelofske2022parallel}.}
    \label{fig:cvrp-qubo}
\end{figure}

The Capacitated Vehicle Routing Problem (CVRP) is an variant of VRP characterized by capacity constrained vehicles, which describes planning tours for vehicles to supply a certain number of passengers (see Figure~\ref{fig:cvrp-qubo} for an illustration). The challenge is exploded complexity with the increasing passenger number.
Feld et al. implemented D-Wave's QA to accelerate the optimization process. In order to solve this problem on a quantum machine, CVRP was mapped to a QUBO \cite{feld2019hybrid}. They investigated different quantum-classic approaches, expound their difficulties in finding good solutions, and finally proposed a hybrid method based on the Two-Phase-Heuristic algorithm \cite{laporte2002classical}.
Irie and their team presented a novel formulation of CVRP using QUBO formulation with time, state, and capacity \cite{irie2019quantum}. The rigorous definition of time enabled to establish various time-related constraints, while the introduction of capacity-qubits ensured model pickup and delivery operations during each vehicle trip.
Azad et al., solved this VRP problem by minimizing corresponding Ising Hamiltonian using a hybrid quantum-classical heuristic called quantum approximate optimization algorithm \cite{azad2022solving}.

\subsection{Quantum Computing for Autonomous Driving} \label{subsec:autonomous}
Quantum computing may be applied to autonomous driving in several key areas, including perception \& sensor fusion, path planning, and cybersecurity \cite{burkacky2020will, larasati2022trends}. Quantum computing could potentially facilitate the coordination and communication between autonomous vehicles, enabling them to share information and make collective decisions for optimal traffic management and safety.

\begin{figure}[ht!]
    \centering
    \subfloat[Classical RPN]
    {\includegraphics[width=0.6\linewidth]{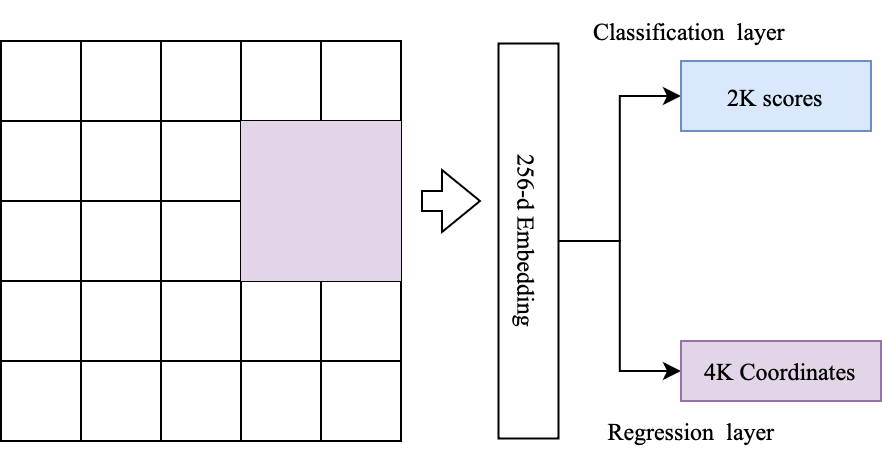}} \\
    \subfloat[Quantum RPN]
    {\includegraphics[width=0.9\linewidth]{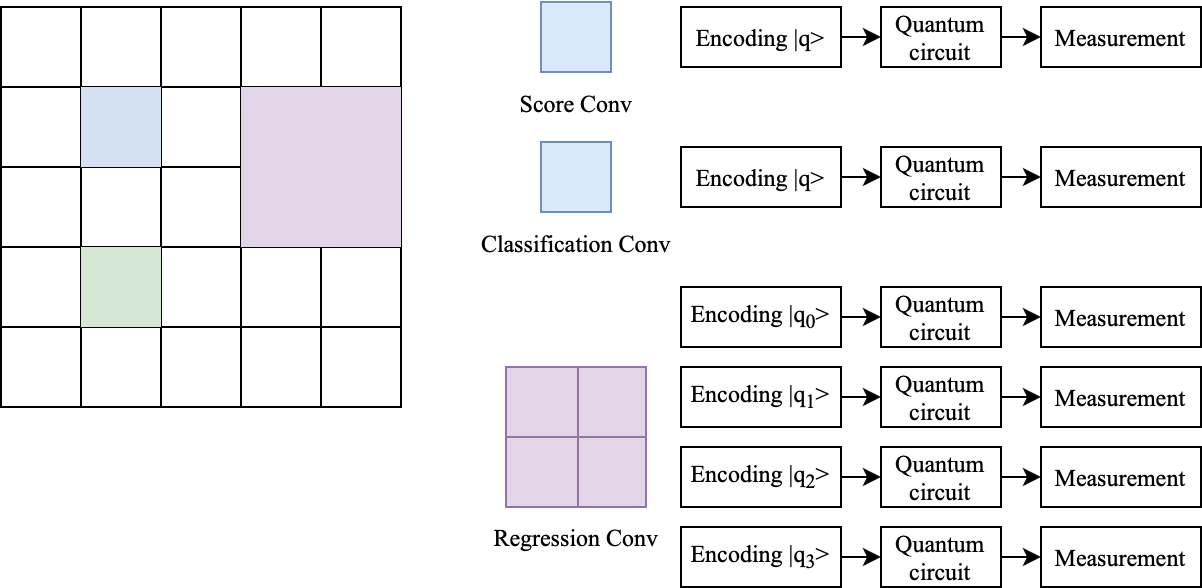}} \\
    \caption[Comparison of Classical and Quantum RPN]{Comparison of Classical and Quantum RPN.}\label{fig:comparison_rpn}
\end{figure}

Perception is a fundamental part for autonomous driving \cite{chen2023milestones}, providing information to following surrounding-object motion prediction and driving planning. Perception algorithms have been well developed based on various sensors and are running on GPU/TPU.
Baek et al., implemented a low-complexity object detection algorithms based quantum convolution neural networks \cite{baek2023fast}. The fast quantum convolution encodes multiple channels into quantum states and achieves quantum speed-ups. The comparisons between classical and quantum Region Proposal Network (RPN) are shown in Figure \ref{fig:comparison_rpn}. However, training the quantum network is challenging due to the limited availability of quantum computing resources and optimization tools. Thus, they utilized the pre-trained detection model as the teacher model to conduct knowledge distillation.
Majumder et al., built a hybrid classical-quantum deep learning model for traffic image classification under adversarial attack\cite{majumder2021hybrid}. The classical part is a classical CNN model for feature extraction. Then the extracted features are processed through the quantum layer composed of various quantum gates.
The vision-based traffic anomaly detection is a challenging research area due to the low frequency of anomalous cases but significant for traffic safety. Amin et al., built a quantum network for video analysis \cite{amin2023detection}. A quantum field is formed by embedding a small portion of input data, e.g., a $2\times2$ rectangle is used to parameterized rotations on ground state qubits. The machine performs a quantum calculation connected with a unitary from the variational quantum circuit \cite{henderson2020quanvolutional}.

Driving behavior analysis  is another important research area for autonomous driving. Sheu deduced macroscopic traffic conditions by creating a quantum-based microscopic model that simulates the immediate decision-making processes of drivers as they navigate past an accident scene in neighboring lanes \cite{sheu2008quantum}. They referred to the theorem of quantum mechanics in optical flows for initial stimulus in driver decision process to explain the motion-related perceptual phenomena while vehicles approach the incident site in adjacent lanes.
Bruza et al., utilized quantum cognition \cite{bruza2015quantum} to replace classical cognition models for analyzing intention of human behavior of the estimated traffic participants and their interaction \cite{song2022research}.

Quantum computing might also benefit the safe and reliable communications cross vehicles \cite{arun2014review}. 
Xiao et al., designed a new security scheme for facilitating vehicular communication by employing a quantum secret sharing (QSS) method \cite{xiao2019efficient}. The proposed QSS scheme could effectively prevent potential adversary and maliciously-behaved vehicles to share the secret besides improving communication efficiency.
In a cloud-supported system environment, Islam et al., combined quantum and classical neural networks to detect an amplitude shift cyberattack \cite{islam2022hybrid}. The classical network is responsible for complex and high-dimensional feature extraction. Its outputs that are in a more informative space are processed by the quantum network.

\section{Quantum Computing Challenges}
Although quantum computing shows promising speedup in specific tasks, it still faces challenges from hardware, software, and implementation perspectives in practical applications \cite{corcoles2019challenges}.

From the hardware perspective \cite{de2021materials}, maintaining the fragile quantum states of qubits is difficult due to decoherence caused by environmental noise and interactions \cite{divincenzo1998decoherence}. Quantum computers, unlike classical computers, are highly sensitive to noise. The delicate quantum state of qubits can be disrupted by even minor disturbances like vibrations or temperature fluctuations, leading to uncontrollable changes in the computer and potential data loss.
Quantum systems are error-prone, requiring robust error correction methods due to above quantum decoherence \cite{roffe2019quantum}. In classical computers, errors are infrequent and typically involve bit flips. However, quantum computers experience errors much more frequently, and these errors can take the form of phase flips, bit flips, or a combination of both. Furthermore, due to the no-cloning theorem in quantum mechanics, replicating qubits for error correction, as is done with classical bits, is impossible. The error correction methods usually protects the quantum information stored in one logical qubit from errors by encoding it into several physical qubits. Thus, the higher the error rate is, the more physical qubits are need.
Building quantum computers with a large number of qubits while maintaining coherence and low error rates remains a major engineering challenge \cite{takeda2019toward}. 

From the software perspective, creating efficient quantum algorithms for specific problems requires new ways of thinking and innovative approaches \cite{fingerhuth2018open}. Quantum programming languages differ significantly from classical languages, which primarily deal with deterministic logic and data manipulation. Quantum languages require a deeper understanding of quantum phenomena like qubits, quantum gates, superposition, and entanglement. Developers need to grasp the underlying physics and quantum algorithms to write effective quantum programs, demanding a significant shift in programming mindset and expertise.
Building the necessary software tools and libraries to support quantum algorithm development and execution is still in its early stages. For example, quantum debugging is difficult \cite{metwalli2022tool}. Traditional debugging methods involve inspecting and troubleshooting a program's state at various points during execution. However, quantum mechanics principles like superposition and entanglement prevent direct observation of a quantum program's intermediate states. The act of measurement collapses the quantum state, making classical debugging techniques inapplicable. This necessitates the development of new debugging tools and strategies specifically for quantum programs.

From the implementation perspective, quantum computers are currently extremely expensive to build and maintain, limiting their accessibility. Many quantum computing technologies require extremely low temperatures, i.e., around close to absolute zero kelvin to maintain the delicate quantum states of their qubits \cite{narasimhachar2015low}. Meanwhile, quantum computers will consume significant amounts of energy to keep an extreme low temperature before advanced progress in superconductor materials, which also raises environmental concerns.

\section{Conclusion}
ITS has significantly benefited from emerging technologies such as machine learning, to process a vast amount of traffic data, optimize traffic efficiency, enhance road safety, and considers sustainability and equity. Quantum computing shows early promise in applications of transportation sector. This paper firstly introduces fundamental knowledge of quantum computing then reviews state-of-the-art quantum computing applications in ITS. By addressing challenges and focusing on future research and development, the full potential of quantum computing could be unlocked in the coming years, leading to safer, more efficient, and sustainable transportation systems.



\bibliographystyle{ieeetr}
\bibliography{reference}

\end{document}